%% file: skeleton.tex
\documentclass[a4paper,11pt]{article}
\usepackage{pos}
\usepackage{cleveref}
\Crefname{equation}{eq.}{eqs.}
\Crefname{section}{sec.}{secs.}
\Crefname{figure}{fig.}{figs.}
\Crefname{appendix}{appendix}{appendices}

\include{shortcuts}

\title{Three-particle scattering
amplitudes from lattice QCD}

\author*[a]{Stephen R. Sharpe}

\affiliation[a]{Physics Department, University of Washington\\
  Seattle, WA 91895-1560, USA}

\emailAdd{srsharpe@uw.edu}

\abstract{I review recent progress in calculating scattering amplitudes and resonance properties involving three particles using results from lattice QCD. The necessary input is the finite-volume spectrum, and the outputs---via solutions of integral equations---are scattering amplitudes that can be continued into the complex plane to search for resonance poles. I describe the outlook for future extensions and applications of this work.
}

\FullConference{The 42nd International Symposium on Lattice Field Theory (LATTICE2025)\\
2-8 November 2025\\
Tata Institute of Fundamental Research, Mumbai, India\\}

\begin{document}
\maketitle

\section{Introduction}\label{sec:intro}
In this talk, I provide a status report on the calculation of scattering amplitudes involving three hadrons from first principles using lattice QCD (LQCD).  I will mainly consider progress since the last dedicated plenary talk at a lattice conference on this topic,\footnote{%
There have also been several plenaries on hadron spectroscopy and interactions in the intervening years that have covered aspects of the present topic~\cite{Mathur:2025,Erben:2025zph,Romero-Lopez:2025veh,Hanlon:2024fjd,Romero-Lopez:2022usb,Horz:2022glt}; see also Jeremy Green's plenary on hadron spectroscopy at this meeting~\cite{Green:2025}.}
given by Akaki Rusetsky in 2019~\cite{Rusetsky:2019gyk}. 
This is a rather specialized and technical area, and I aim to give an overview of the important issues while largely avoiding details.

There are several very interesting adjacent topics that I will not cover due to lack of time and space.
These are (i) the development and application of methods to calculate matrix elements involving resonances (see talks by Ortega-Gama and Sakthivasan, as well as the 2024 plenary by Erben~\cite{Erben:2025zph});
(ii)
the calculation of electroweak decay amplitudes (see talks by Erben and Lundstrum, as well as 2023 plenary by Leskovec~\cite{Leskovec:2024pzb});
and (iii)
the development of methods for calculating properties of inclusive decays
(talks by Abbott, Elgaziari, Fields, Garofalo, Gavriel, Margari, Moriandi, Pino Rubio, and Zimmermann,
as well as the 2022 plenary by Bulava~\cite{Bulava:2023mjc}).
I will also not cover recent beautiful work 
that determines the ground state energy of multiparticle systems---up to 6144 pions---and from that extracts information about multiparticle interactions and the finite-density equation of state~\cite{Abbott:2023coj,Abbott:2024vhj}.

Returning to the central topic, one might wonder what we hope to learn from a calculation of three-particle amplitudes.
In my view, there are three major motivations for this effort: first, to predict the properties of resonances;
second, to have first-principles calculations of three-body interactions;
and, third, as a stepping stone to the calculation of electroweak decay amplitudes such as $D\to\pi\pi$.
Let me briefly expand on each of these in turn.

Concerning spectroscopy, 
 LQCD has successfully carried out the benchmark calculation of the masses of hadrons that are stable under strong interactions,\footnote{%
 Which I will simply refer to as ``stable'' in the remainder of this talk.}
  and also of the lowest lying resonances where two-particle decays dominate.
 See Jeremy Green's plenary talk at this meeting for a summary of the present status~\cite{Green:2025}.
For example, the properties of the $\rho$ resonance, which predominantly decays to two pions, have been successfully determined using the two-particle finite-volume formalism that was introduced long ago by L\"uscher~\cite{Luscher:1986pf,Luscher:1991n1,Luscher:1991n2}
 and since generalized by many authors.
 I will refer to this formalism as the two-particle quantization condition, or QC2 for short. 
 However, most resonances have decay channels involving three or more stable hadrons.
Notable low-lying examples are the $\omega(780)$ and $\pi(1300)$,
 both of which decay predominantly to three pions,
since G-parity forbids decays to even numbers of pions in isosymmetric QCD.
Also of great interest is the Roper resonance, $N(1440)$, which decays both to $N\pi$ and $N\pi\pi$,
and remains poorly understood in the quark model.

To determine the properties of such resonances, which are not asymptotic states of QCD, one needs to calculate the corresponding three-particle scattering amplitudes, e.g. $\mathcal M(3\pi \to 3\pi)$,
and determine the position of complex poles in such amplitudes.
This motivation has become more urgent in the last decade or more, because of the experimental discovery of a cornucopia of new resonances, many of them quark-model exotics, e.g. the doubly charmed tetraquark, $T_{cc}(3875)^+$~\cite{LHCb:2021vvq,LHCb:2021auc}. This has breathed new life into the topic of spectroscopy, showing that a strongly-interacting theory such as QCD has many qualitative properties that we do not understand. Many of the new resonances decay to three or more stable particles, e.g. $T_{cc}^+ \to D D^* \to D D\pi$, and thus control over the corresponding three-particle scattering amplitudes is needed to predict their properties.

Turning to the second motivation, the aim is to use LQCD to predict the short-distance three-particle interactions that are needed to study dense matter. Here the ultimate goal is the three-nucleon interaction, the lack of knowledge of which is one of the uncertainties in the prediction of the properties of large nuclei and of the neutron star equation of state~\cite{Hoppe:2019uyw,Drischler:2020hwi,Machleidt:2024bwl}.

The final motivation---the calculation of electroweak decay amplitudes---is part of the highly successful, decades-long effort by LQCD to test the standard model using flavor physics. See the plenary talk by Gregorio Herdo\'iza for a status report on that endeavor. The methods needed to obtain decay amplitudes such as $\mathcal A(K\to2\pi)$ 
and  $\mathcal A(K\to3\pi)$
are closely related to those needed to determine scattering amplitudes. 
A long-term dream is to extend
these methods to the calculation of $D$-decay amplitudes, e.g. $\mathcal A(D\to \pi\pi)$
in which CP violation has been measured~\cite{LHCb:2019hro,LHCb:2022lry}.

In the short term, the spectroscopy applications will be dominant, and, in that regard, I cannot resist adding a personal motivation to the list above:
It is just plain fun to be able to use the femtoworld of LQCD to (indirectly) carry out $3\to3$ scattering ``experiments'', experiments that are inaccessible to direct measurement.
This includes the possibility of using LQCD to study properties as a function of quark mass, allowing one to apply the three-particle formalism in a heavier-than-physical mass regime where higher-particle interactions are less important~\cite{Mai:2021lwb}.
Such ``experiments'' provide nice examples where LQCD can provide complementary information to experiment such that, hopefully, the mysteries of strong-interaction spectroscopy can be disentangled.

One question that the non-expert may well have is this: Why three?
What about processes involving more than three particles?
After all, many resonances have decay channels involving four or more particles.
There is, indeed, much interest in extending the work to more than three particles 
(a point that I return to at the end of this talk),
but the results available to date have concerned three particles.
It has turned out that the three-particle sector has been richer and more complicated than I, at least, expected.

In the remainder of this talk, I first describe the underlying theoretical issue and the status of the formalism that resolves it---which I will loosely refer to as the three-particle quantization condition or QC3---and then summarize several recent and ongoing applications of the formalism that use LQCD to predict amplitudes and resonance properties. I close with a rather lengthy outlook.

\section{Status of formalism}

The situation facing a LQCD theorist desirous of calculating scattering amplitudes is this:
LQCD is very effective at determining the spectrum of finite-volume states in spatial boxes of size $2-10\,$fm, using Euclidean correlation functions, but the scattering amplitudes that we want are infinite-volume quantities involving in and out states. How are the spectrum and amplitudes related?
In this regard, it is important to keep in mind that the finite-volume energies are physical quantities (after taking the continuum limit), so that we are asking for a relation between one set of physical quantities and another.
And it is plausible that, for a box large enough that the hadrons do not significantly overlap, 
the finite-volume energies depend on the scattering amplitudes. The question is how to extract this information.

L\"uscher solved this for the case of two particles~\cite{Luscher:1986pf,Luscher:1991n1,Luscher:1991n2}.
This works as follows (using a notation convenient for the generalization to three particles). Given the two-particle K matrix, $\mathcal K_2$, an infinite-volume quantity, 
the finite-volume spectrum is given by the values of $E$ that solve the QC2
\begin{equation}
\det_{\ell m} \left( \mathcal K_2(E^*) + F_2(E,\boldsymbol P,L)^{-1}\right) = 0\,.
\label{eq:QC2}
\end{equation}
Here $\boldsymbol P$ is the (quantized) total momentum in the box, which I assume is cubic with length $L$.\footnote{%
The generalization to other spatial box shapes is straightforward~\cite{Lee:2017igf,Culver:2019qtx}.}
Thus the center of momentum (CM) energy is given by $E^*=\sqrt{E^2-\boldsymbol P^2}$, and it is this that the two-particle K matrix depends on.
Both quantities in \Cref{eq:QC2} are matrices in $\{\ell,m,x\}$ space, where $\ell$ is the relative angular momentum of the two particles in the CM frame, $m$ is the corresponding azimuthal quantum number,
and the mystery index $x$ includes the effects of spin and multiple two-particle channels,
and takes values that depend on the system~\cite{Briceno:2014oea}.
 $\mathcal K_2$ is diagonal in $\ell$ and $m$, whereas the purely kinematical function $F_2$ (the ``L\"uscher zeta function'') is, in general, not diagonal, reflecting the violation of rotational symmetry by the finite box.\footnote{%
Explicit expressions for $F_2$, and the relation of $\mathcal K_2$ to phase shifts, are given, 
e.g., in Ref.~\cite{\DRS}.
}

The QC2 accounts for power-law dependence on $1/L$, and holds up to corrections that fall as $\exp(-M_\pi L)$ (up to powers of $L$), where $M_\pi$ enters as it is the lightest scale in a hadronic system. For $M_\pi L \gtrsim 4$, such exponentially-suppressed corrections are at the percent level or smaller. They can be studied in various ways, e.g. by using a range of values of $L$ in the simulations. 

A potentially more important source of error comes from the fact that, to use the QC2, one must truncate the angular momentum index $\ell$ in order to have finite matrices.
This is achieved formally by setting $\mathcal K_2$ to zero for $\ell > \ell_{\rm max}$.
Near to threshold, barrier factors lead to $\mathcal K_2 \propto q^{2\ell}$, where $q$ is the relative momentum of the two particles in their center of momentum frame,  so such a truncation is reasonable.
At higher energies it is a less controlled approximation, though it is one that is commonly used also in analysis of experimental data.

The QC2 is strictly valid only in the range of $E^*$ for which there are only two-particle channels, i.e. up to the first inelastic channel (e.g. $E^*=4 M_\pi$ in the case of the $\rho$, assuming G parity).
The formalism for multiple two-particle channels, with particles of arbitrary spin, is known~\cite{Briceno:2014oea} and has been implemented in several cases.

The workflow that has been typically used in practice involves assuming a parametrization of $\mathcal K_2$, and then adjusting the parameters so as to give the best fit of the spectrum predicted by \Cref{eq:QC2} to that obtained from the simulation, accounting for statistical errors. Ideally one first extrapolates the spectrum to the continuum limit, although to date that has rarely been done in practice. Simultaneous fits to several values of $L$ are preferable.

The K matrix is related to the scattering amplitude by an algebraic equation
\begin{equation}
\mathcal M_2 = \mathcal K_2 \frac1{1 - i \rho \mathcal K_2}\,,
\label{eq:KtoM2}
\end{equation}
where $\rho$ is the phase-space matrix (given explicitly in, e.g., Ref.~\cite{\DRS};
one should beware, however, that different definitions of K matrices are used in literature).
Given that $\mathcal K_2$ is an hermitian matrix, this relation ensures that $\mathcal M_2$ leads to a unitary S matrix, and, in particular, includes the phases induced by initial- and final-state interactions.
Using this result, and the parametrization of $\mathcal K_2$, one can then analytically continue onto the second Riemann sheet (or sheets, in the case of multiple two-particle channels), and determine resonance parameters.

The application of this workflow for the QC2 is by now quite standard, and, in some cases, approaching ``FLAG readiness'', i.e. with all errors controlled.
See Jeremy Green's talk for an update~\cite{Green:2025}.

I now turn to the three-particle system. The formalism was developed first for three identical scalar particles with a G-parity-like symmetry, following three different approaches. The first used an all orders diagrammatic analysis in a relativistic effective field theory (EFT) describing generic interactions between the particles, and is dubbed the RFT approach~\cite{\HSQCa,\HSQCb}. This approach makes no assumption about the form of the interactions, nor does it rely on a power counting.
The second used nonrelativistic (NR) EFT, and led to a much simpler derivation~\cite{\Akakia,\Akakib}, although the three-particle interaction was initially restricted to the leading order terms in the NR power counting.
Subsequent work has shown how to relativize this approach~\cite{Muller:2021uur}.
The third approach used a unitary representation of the three-particle amplitude developed in the infinite volume~\cite{\Maiisobar}, and from that developed a finite-volume quantization condition~\cite{\MD}.
It is denoted the ``finite-volume unitarity'' or FVU approach.

Much follow-up work tested the formalisms~\cite{\HSPT,\HSTH,\HSBS,\SPT,\AkakiTHa}, 
and made initial numerical investigations, largely, though not entirely, in toy models~\cite{\BHSnum,\MDpi,\Akakinum,\dwave,Culver:2019qtx,\largera}.
Roughly speaking, this was the situation in 2019, as discussed in Rusetsky's review~\cite{Rusetsky:2019gyk}.
That review provides an overview of the history, and a derivation using the NREFT approach.
The status was also reviewed around the same time in Ref.~\cite{\HSrev}, largely from the RFT perspective,
but also studying the relation between formalisms.

While there have been many extensions of the formalism in the intervening years, the essential forms of the
results have not changed, and thus I will simply quote the results, choosing the RFT approach, and point to the above-mentioned reviews and the original papers for the details.\footnote{%
See also the more recent reviews in Refs.~\cite{\MDRrev,Romero-Lopez:2021zdo}.
}
 I will also list the advances that have been made, and return in the outlook to what I see as the remaining unresolved issues.

The workflow for determining the three-particle scattering amplitude consists, in all approaches, of two steps.
I will describe these for the case in which there are no transitions between two and three particles, e.g. three pions in isosymmetric QCD.
In the first step, one fits the finite-volume energies for two and three-particles (which, by assumption, can be calculated separately) to the predictions of the two- and three-particle quantization conditions, respectively (QC2 and QC3), and thus determines (the parametrized forms) of the two- and three-particle K matrices.
In the second step, one solves integral equations in which the K matrices are kernels, the output of which is the three-particle scattering amplitude, $\cM_3$.

The QC3 appearing in the first step has the form
\begin{equation}
\det_{k \ell m x} \left( \mathcal K_{\rm df,3}(E^*) + F_3(E,\boldsymbol P,L,\cK_2)^{-1}\right) = 0\,,
\label{eq:QC3}
\end{equation}
where (for explicit expressions see, e.g., Ref.~\cite{\DRS}; note also that in some RFT works so-called asymmetric forms of $F_3$ and $\Kdf$ are used~\cite{Blanton:2020gha,Jackura:2022gib,Briceno:2024ehy,Jackura:2025wbw,Briceno:2025yuq})
\begin{equation}
F_3 = \frac1{2\omega L^3} \left[ \frac{F}3 - F \frac1{\cK_{2,L}^{-1} + F + G} F \right]\,.
\label{eq:F3}
\end{equation}
As for the QC2, solutions to the QC3 give the values of $E$ at which there are finite-volume energies for the given K matrices. All quantities in these equations are matrices in a space with indices $\{k,\ell, m, x\}$---a larger space than in the QC2.\footnote{%
I stress that all three approaches to the three-particle formalism use matrices with the same index space.}
The formalism separates the three particles into a spectator, with momentum $\boldsymbol k$,
and the remaining pair. The index $k$ is a shorthand for the finite-volume momentum of the spectator, 
$\boldsymbol k \in (2\pi/L) \mathbb Z^3$. The indices $\ell,m$ represent the relative angular momentum of the particles in the pair in their CM frame. Finally, the index $x$ plays a similar role to that in the QC2, and accounts for the effects of flavor,  spin, and the possibility of multiple three-particle channels. It always takes a finite number of values, and is absent in the case of three identical particles.

A key feature of the QC3 is that it involves a cutoff function (buried inside the various matrices) that cuts off the sum over $\boldsymbol k$, such that the invariant mass of the pair cannot go too far below threshold. 
This is needed to avoid subthreshold singularities in the K matrices (which can lead to uncontrolled power-law volume dependence), and ensures that the matrices have finite dimension 
(assuming, as for the QC2, a truncation in $\ell$).

The matrices $F$ and $G$ are known kinematic quantities depending on $E, \boldsymbol P$, and $L$.
They give rise to the power-law volume dependence in the spectrum,
and are associated with on-shell intermediate states of three particles.
$F$ is a simple generalization of $F_2$, in which the pair plays the role of the two-particle system in the QC2, with the spectator spectating.
$G$ is a three-particle quantity, associated with processes in which the spectator particle changes.
The matrix $\omega$ is a simple kinematic quantity containing the energy of the spectator.

Although the QC3 is superficially similar to the QC2, this hides a number of significant differences.
In addition to the larger matrix space, a key difference is that $F_3$ is not a purely kinematic function, unlike $F_2$ in the QC2. In particular, it depends on the two-particle K matrix, packaged into $\mathcal K_{2,L}$.
Thus $F_3$ accounts for the effects of two-particle interactions on the finite-volume spectrum,
which dominate over those of three-particle interactions at large $L$.
The latter are incorporated through the three-particle K matrix, $\cK_{\rm df,3}$.

$\cK_{\rm df,3}$ is an hermitian, infinite-volume quantity (which can be chosen real in single-channel systems),
and represents the short-distance part of the three-particle interaction.
Like $\cM_3$, it depends on $E^*$ and the seven other variables describing $3\to3$ scattering.
However, unlike $\cM_3$, it depends on the cutoff function and is thus unphysical. 
In the QC3, the cutoff dependence cancels that buried in $F$, $G$ and $\cK_{2,L}$, 
so that the spectrum itself is physical.
The subscript ``df'' on $\cK_{\rm df,3}$ stands for ``divergence-free'', 
and indicates that the well-known divergences in $\cM_3$ due to one-particle exchange (OPE) and higher-order exchange processes~\cite{Rubin:1966zz} are absent in $\cK_{\rm df,3}$.

As for the QC2, the QC3 is valid only up to exponentially-suppressed corrections, and also only up to the first relevant inelastic threshold, i.e. the value of $E^*$ above which on-shell intermediate states with more than three-particles appear. For example, for the three-pion system, $E^*_{\rm max} = 5 M_\pi$.

\begin{figure}[h!]
\centering
\includegraphics[width=\textwidth]{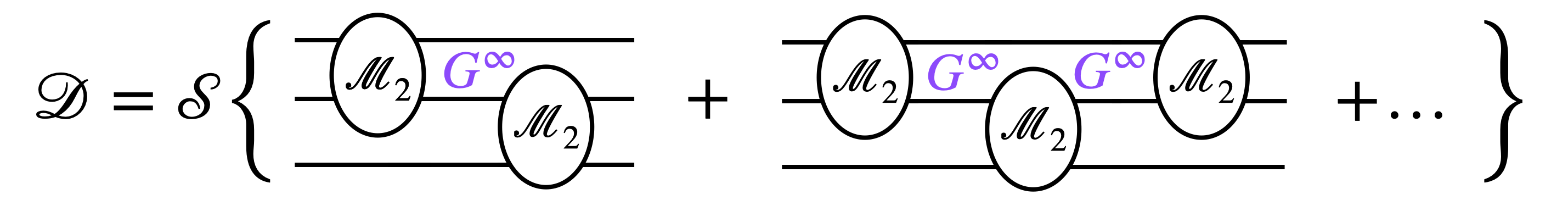}
\includegraphics[width=\textwidth]{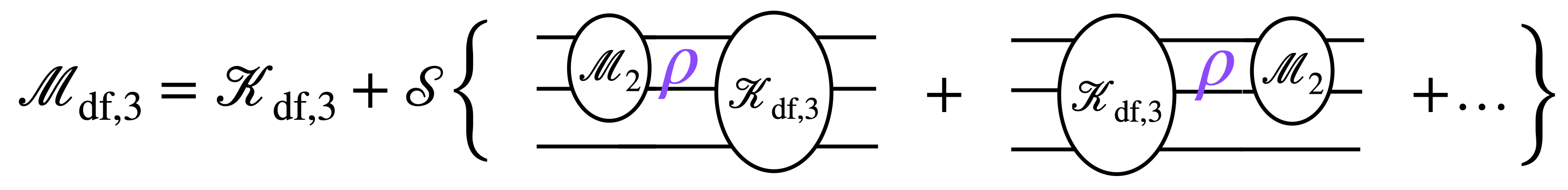}
\caption{Schematic form of the integral equations for $\cD$ and $\cM_{\rm df,3}$. Notation is explained in the text.}
\label{fig:inteq}
\end{figure}

Now I turn to the second step in the three-particle formalism: solving the integral equations.
Here I will be even more schematic.
The high-level form of the RFT result is~\cite{\HSQCb}
\begin{equation}
\cM_3 = \cD(\cK_2) + \cM_{\rm df,3}(\cK_2,\Kdf)\,,
\label{eq:inteq}
\end{equation}
where the equations for $\cD$ and $\cM_{\rm df,3}$ are illustrated in \Cref{fig:inteq}.
Each involves an infinite series of terms that can be recast as a set of nested integral equations.
Explicit forms are given in the references below.
$\cD$ sums the effects of multiple OPEs between two-particle interactions. It involves $\cM_2$
(related to $\cK_2$ by \Cref{eq:KtoM2}) and a modified exchange propagator, $G^\infty$, that contains the pole
of the standard Feynman propagator, but also the cutoff function and other modifications.
The operator $\cS$ symmetrizes the expression over choices of the spectator.

The divergences in $\cM_3$ are contained in $\cD$.
The remaining, divergence-free part $\cM_{\rm df,3}$, adds in all terms containing at least one factor of $\cK_{\rm df,3}$. Here $\rho$ indicates a phase-space factor, which again includes the subthreshold cutoff function.
One can think of the various ``decorations'' of $\cK_{\rm df,3}$ as inserting the initial and final-state interactions (between two and three particles) such that $\cM_3$ satisfies $s$-channel unitarity.
An explicit demonstration of this is given in Ref.~\cite{\BHSSU}.

Over the last six years or so, there has been tremendous progress in three directions: 
extending the formalism to essentially all three-particle systems of interest in QCD;
developing methods for solving the integral equations;
and applying the complete formalism to a range of interesting and phenomenologically relevant systems.
I have summarized this progress in \Cref{tab:formalism,tab:inteqs,tab:QC3apps}.
Other advances include the simplification of the RFT derivation using time-ordered perturbation theory~\cite{\BSQC},
the demonstration of the equivalence of the RFT and FVU formalisms both for the QC3~\cite{\BSequiv} 
and the integral equations~\cite{Jackura:2019bmu};
the development of an alternative approach to the derivation of the formalism based on unitarity~\cite{Jackura:2022gib}; 
the calculation of $\cK_{\rm df,3}$ for three pions of arbitrary isospin at
next-to-leading order (NLO) in chiral perturbation theory 
(ChPT)~\cite{Baeza-Ballesteros:2023ljl,Baeza-Ballesteros:2024mii};
and the extension of threshold expansions within the NREFT approach~\cite{Muller:2020vtt,Bubna:2023oxo}.

\begin{table}[htp]
\begin{center}
\begin{tabular}{c|c|c}
System & Example & References \\
\hline
$2+3$ identical scalars & $\phi^3$ theory & \cite{\BHSQC}(RFT) \\
$3\pi$ arbitrary isospin & $\pi^+\pi^0 \pi^- \leftrightarrow \omega$ & \cite{\isospin}(RFT), \cite{\Maiaone,Feng:2024wyg,Yan:2024gwp,Yan:2025mdm}(FVU)\\
3 nondegenerate scalars & $\pi^- D^0 D_s^+$ & \cite{\BSnondegen}(RFT) \\
2 degenerate, identical scalars $+1$ & $\pi^+\pi^+ K^+$, $\pi^+ K^+K^+$ & \cite{\BStwoplusone}(RFT), \cite{Feng:2024wyg}(FVU)\\
3 identical spin-1/2 particles & $3n$ & \cite{\threeneutron}(RFT) \\
$DD\pi$ all isospins & $T_{cc}^+$ & \cite{\tetraquark}(RFT)\\
Multiple 3-particle channels & $K^-K^+\pi^0 \leftrightarrow \pi^0\pi^0\eta$ & \cite{\multichannel}(RFT)\\
$N\pi\pi$ I=5/2 & $p p \pi^+ \leftrightarrow \Delta^{++} p$ & \cite{\Npp}(RFT)
\end{tabular}
\end{center}
\vspace{-0.1in}
\caption{Status of formalism beyond that for three identical scalars, all for isosymmetric QCD.}
\label{tab:formalism}
\end{table}%

\begin{table}[htp]
\begin{center}
\begin{tabular}{c|c}
System & References \\
\hline
$3\pi$, $I=3$& \cite{Hansen:2020otl}(RFT) \\
$3\phi \leftrightarrow [\phi\phi]+\phi \leftrightarrow [\phi\phi\phi] $ & \cite{Jackura:2020bsk,Dawid:2021fxd,Dawid:2023jrj,Dawid:2023kxu}(RFT)
\\
$3\phi_1 \leftrightarrow \phi_2$ & \cite{Garofalo:2022pux}(FVU \& RFT)
\\
$3\pi$, $I=0-2$ & \cite{\Maiaone,Yan:2024gwp,Yan:2025mdm}(FVU), \cite{Jackura:2023qtp, Briceno:2024ehy, Jackura:2025wbw,Briceno:2025yuq}(RFT)
\\
$3\pi^+$, $\pi^+\pi^+ K^+$, $\pi^+K^+K^+$, $3K^+$ & \cite{Dawid:2025zxc,Dawid:2025doq}(RFT)
\\
$\pi^0D^0D^+ \leftrightarrow \pi^- D^+ D^+ \leftrightarrow T_{cc}^+ $ & \cite{\DRS,Dawid:2025wsn}(RFT)
\end{tabular}
\end{center}
\vspace{-0.1in}
\caption{Status of solutions to integral equations. $[\phi\phi]$ and $[\phi\phi\phi]$ indicate, respectively, a bound-state dimer and trimer composed of scalars $\phi$. All cases other than the $3\phi$ system involve end-to-end applications of the formalism, starting with spectra obtained from lattice simulations.}
\label{tab:inteqs}
\end{table}%

\begin{table}[htp]
\begin{center}
\begin{tabular}{c|c}
System & References \\
\hline
$3\phi\leftrightarrow [\phi\phi]+\phi$ & \cite{Romero-Lopez:2019qrt}(RFT) \\
$3\pi$, $I=3$
& \cite{\HHanal,Fischer:2020jzp}(RFT), \cite{Mai:2018djl,Mai:2019fba, Culver:2019vvu, Brett:2021wyd}(FVU) \\
$3 K^-$ & \cite{Alexandru:2020xqf}(FVU)\\
$3\pi^+$, $\pi^+\pi^+ K^+$, $\pi^+K^+K^+$, $3K^+$ & \cite{Blanton:2021eyf,Blanton:2021llb,Draper:2023boj}(RFT)\\
\end{tabular}
\end{center}
\vspace{-0.1in}
\caption{Applications of the QC3 that determine $\cK_{\rm df,3}$ (or set it to zero) but do not solve integral equations.}
\label{tab:QC3apps}
\end{table}%

The result of this progress is that a number of nontrivial and phenomenologically interesting applications have appeared---and these will be my focus for the remainder of this talk.\footnote{%
I will not discuss Ref.~\cite{Briceno:2025yuq} in detail, as it appeared only a few days before the conference began.}

\section{Recent applications of the three-particle formalism}\label{sec:apps}

\subsection{Three-particle amplitudes involving pions and kaons at physical quark masses}

The recent work of Ref.~\cite{Dawid:2025zxc,Dawid:2025doq} presents the first determination of a three-particle scattering amplitude for physical quark masses (albeit in isosymmetric QCD).
This is a complete application of the formalism---from the spectrum to scattering amplitudes.
It considers a relatively simple system---pions and kaons at maximal isospin---for which the dominant $s$-wave interactions are weakly repulsive, and two- or three-particle bound states are absent.
It is not a FLAG-ready calculation, since it involves only a single lattice spacing ($a\approx 0.063\;$fm) and a single volume. Nevertheless, by combining with previous results from higher quark masses~\cite{Blanton:2021llb,Draper:2023boj}, 
a detailed comparison with ChPT is possible.

The calculation is done on the E250 CLS ensemble~\cite{Bruno:2014jqa},
which was generated with nonperturbatively $\cO(a)$-improved Wilson fermions and the tree-level
$\cO(a^2)$-improved Lüscher-Weisz gauge action. The lattice size is $96^3\times 192$, 
with $a\simeq 0.063\;$fm, such that $M_\pi L = 4.05$. 
Propagators are computed using the stochastic Laplacian-Heaviside method~\cite{Morningstar:2011ka}.
A large operator basis is used, with levels determined using the GEVP method~\cite{Luscher:1990ck}.
For further details see Ref.~\cite{Dawid:2025doq}.

Using this setup, the $2\pi^+$, $\pi^+K^+$, $K^+K^+$, $3\pi^+$, $\pi^+\pi^+K^+$, $\pi^+K^+K^+$, 
and $3K^+$ spectra are determined in nine frames, for all contributing irreducible representations (irreps).
In each channel, levels are obtained up to and beyond the corresponding inelastic energy (e.g. $E^*=5M_\pi$ for
$3\pi^+$, and $E^*=3 M_\pi+M_K$ for $\pi^+\pi^+K^+$).  As an example of the extent of the calculation, 
about 130 $\pi^+ K^+ K^+$ levels are determined. Plots of the spectrum are shown in Ref.~\cite{Dawid:2025doq}.
It is found that the number of interacting and noninteracting levels matches,
with the relative shifts being almost all positive and, in most cases, of high statistical significance.

For each three-particle system, simultaneous correlated fits are performed to the three-particle levels along with those for the corresponding two-particle subchannels, e.g. to $\pi^+ K^+$ and $2K^+$ for the $\pi^+K^+K^+$ system, considering only levels up to the appropriate inelastic thresholds. In these fits, the K matrices are parametrized by truncated threshold expansions, i.e. the effective-range expansion for $\cK_2$ and its generalization for $\cK_{\rm df,3}$~\cite{\dwave,\implement}. 
The most challenging fit is that for $\pi^+ K^+ K^+$, where there are 40 $\pi K$, 25 $KK$,
and 50 $\pi KK$ levels to be fit (115 in total), and 10 parameters. In this case, the fit has $\chi^2/{\rm dof}=1.95$,
but for other systems the fit is better (e.g. $\chi^2/{\rm dof}=1.14$ for $2\pi+3\pi$).

\begin{figure}[h!]
\centering
\includegraphics[width=\textwidth]{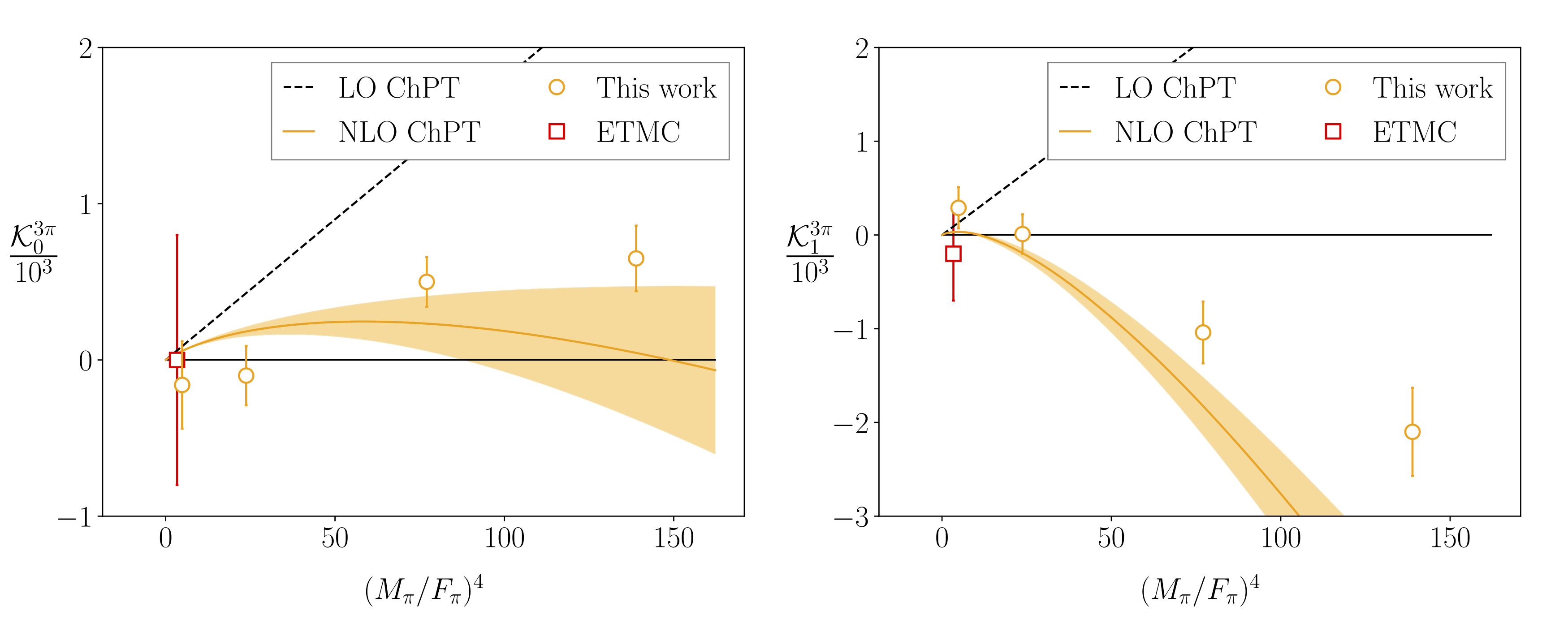}
\caption{Comparison of the leading two terms in $\cK_{\rm df,3}$ for $3\pi^+$ scattering with ChPT (from Ref.~\cite{Dawid:2025doq}); ``ETMC'' refers to Ref.~\cite{Fischer:2020jzp}). NLO ChPT bands are obtained using standard values and errors of low-energy coefficients, see Ref.~\cite{Dawid:2025doq}.
The lowest $M_\pi$ value corresponds to physical quark masses.}
\label{fig:Kdf3pi}
\end{figure}

From the many results in this study, I have selected two of particular interest.
The first concerns the extraction of $\cK_{\rm df,3}$ using the QC3: Is this possible given that
three-body effects on the spectrum are suppressed by $\sim1/L^3$ compared to two-body effects?
In \Cref{fig:Kdf3pi}, I show the comparison between 
the coefficients of the two leading terms in the threshold expansion for 
$\cK_{\rm df,3}$ for $3\pi^+$ scattering and ChPT predictions.
We see that $\cK_{\rm df,3}$ is obtained with relatively small errors at physical masses.
The central value is consistent with zero for physical masses 
(although nonzero results are obtained for $3K^+$ scattering),
but this is consistent with ChPT.
Nonzero results are obtained at higher pion masses, and show an interesting chiral dependence.
The leading term (left panel), in which $\cK_{\rm df,3}$ is simply an energy- and angle-independent constant, 
does not match the predictions of LO ChPT~\cite{\HHanal}, 
but the NLO corrections from Ref.~\cite{Baeza-Ballesteros:2023ljl} bring it into consistency.\footnote{%
As noted earlier, $\cK_{\rm df,3}$ depends on the form of the cutoff function. This is found explicitly in ChPT, where the dependence enters at NLO. In the figure, the same cutoff is used for both the QC3 and ChPT results.}
For the second term (right panel), which incorporates linear dependence on $E^{*2}$ but is otherwise isotropic,
the lattice results have a slope of opposite sign to the LO prediction, but the very large NLO corrections lead to a sign flip in the predictions. Clearly the NLO prediction is not reliable, but the large size of the corrections indicates that there is no inconsistency between the lattice results and ChPT. Only a NNLO calculation can settle the issue.

\begin{figure}[h!]
\centering
\includegraphics[width=0.45\textwidth]{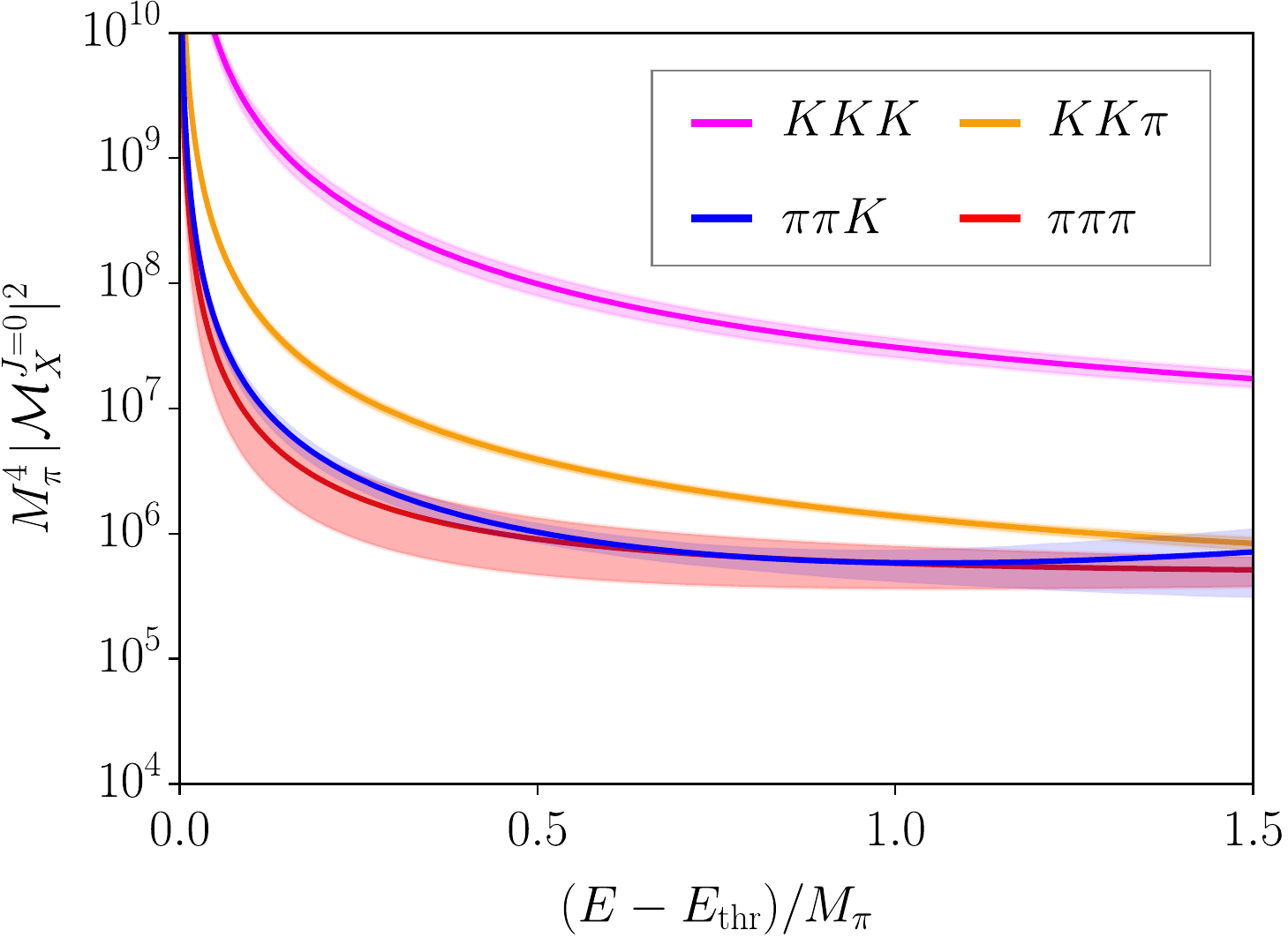}
\includegraphics[width=0.5\textwidth]{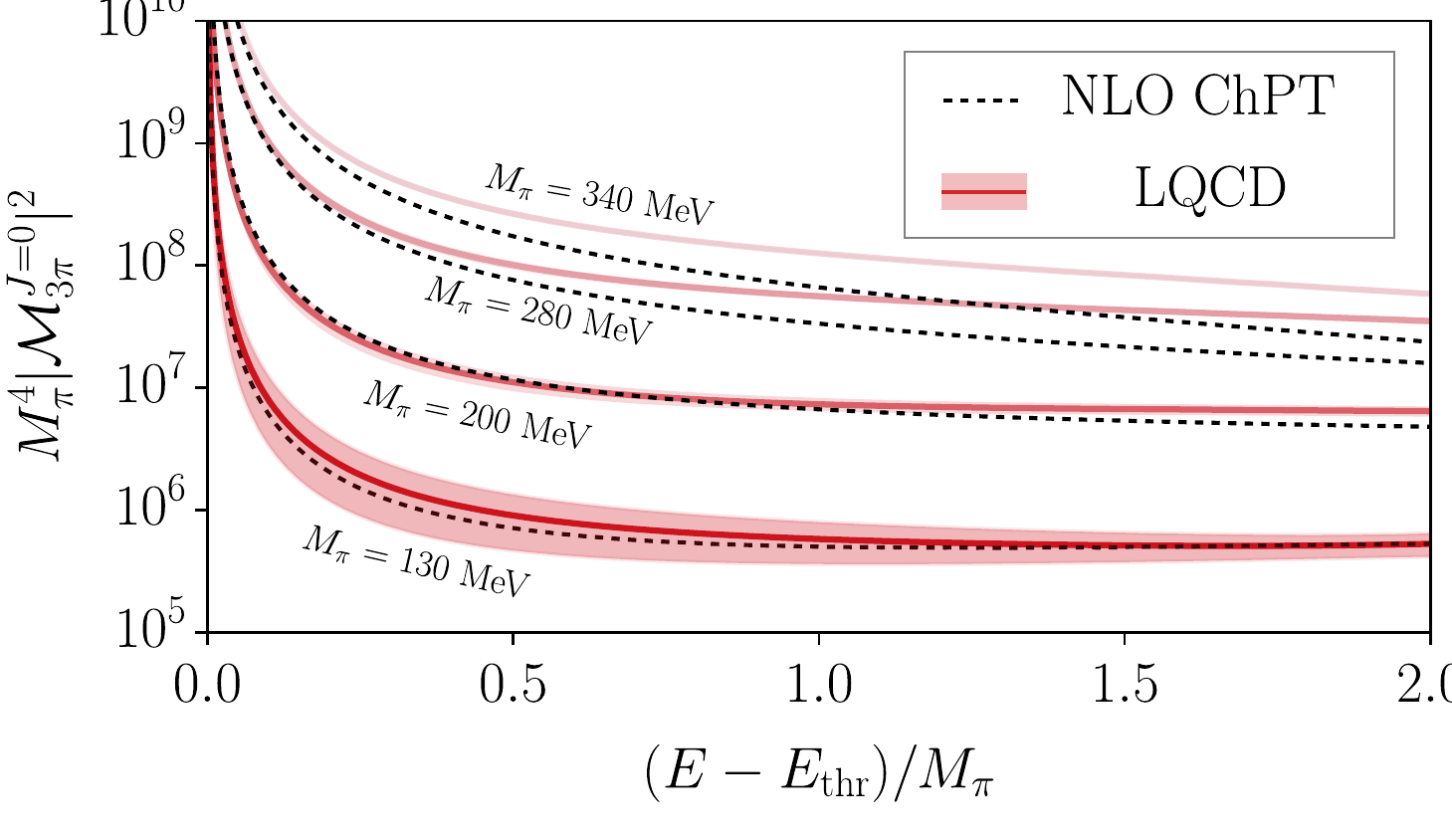}
\caption{Energy dependence of $\cM_3$ with $J^P=0^-$ with incoming and outgoing particles in the equilateral configuration (from Ref.~\cite{Dawid:2025doq}). $E$ is the CM frame energy, $E_{\rm thr}$ is the CM energy at threshold. Left panel: comparing different processes at physical quark masses.
Right panel: comparing $\cM_{3\pi}$ at different quark masses with the NLO ChPT prediction.}
\label{fig:M3}
\end{figure}

Using $\cK_2$ and $\Kdf$ as input, the integral equations for $\cM_3$ have been solved after projection onto definite
total angular momentum and parity.
The leading $J^P=0^-$ amplitude, as well as subleading amplitudes with $J^P=1^+$ and $2^-$, have been
determined.
$\cM_3$ depends on the total energy $E=\sqrt{s}$ and seven other variables.
Examples of the dependence on $E$ for the $J^P=0^-$ amplitude are shown in \Cref{fig:M3}.
Here the kinematics is simplified by choosing the initial and final momenta to lie in an equilateral triangle,
in which case, for $J^P=0^-$, the amplitude depends only on $E$.
We see the divergence of the amplitudes as $E$ approaches the threshold, and (in the left panel) the expected hierarchy in which amplitudes generally decrease as more pions are involved.
The right panel compares $\cM_{3\pi}$ to NLO ChPT, where we see good agreement for physical masses and poorer agreement as $M_\pi$ increases---again as expected.

\subsection{First results for the $\pi(1300)$ resonance}

Another benchmark has recently been achieved by the determination of the parameters of resonances that decay predominantly to three particles. Last year Ref.~\cite{Yan:2024gwp} considered the $\omega(780)$,
while recently Ref.~\cite{Yan:2025mdm} studied the $\pi(1300)$ (see talk by Maxim Mai).
Both use the FVU approach, and are end-to-end calculations from spectra to amplitudes to resonance parameters.
Both use LQCD results at heavier-than-physical quark masses, and extrapolate to physical masses using a ChPT-based model. I would classify the $\omega$ study as exploratory, since it involved only two well-determined energy levels at the pion mass for which the $\omega$ is a resonance,
while the $\pi(1300)$ result is major step forward.

The $\pi(1300)$ is the lightest excited state with the quantum numbers of the pion, 
and has been observed only in the three-pion decay channel~\cite{ParticleDataGroup:2024cfk}.
(G parity forbids decays to even number of pions.)
Its mass ($1300\pm 100\;$MeV) and width ($200$ to $600\;$MeV) are poorly determined.
It is thus a clear target for multiparticle LQCD calculations.
However, it lies well above the inelastic threshold for the QC3, which is at $5M_\pi \approx 700\;$MeV,
making a direct application of the three-particle methodology problematic at physical quark masses.
On the other hand, the QC3 becomes applicable once the pion mass is raised to about $300\;$MeV.

The work of Ref.~\cite{Yan:2025mdm} uses ensembles from the Chinese LQCD collaboration~\cite{CLQCD:2023sdb},
generated with tadpole-improved Symanzik gauge and Clover fermion actions at $a\approx 0.077\;$fm.
Results are obtained for both $M_\pi\approx 305\;$MeV and $210\;$MeV, 
on two lattice sizes, which have $M_\pi L = 3.8$ and $5.7$ for the larger pion mass,
and $M_\pi L = 2.7$ and $3.9$ for the smaller.
Many operators are used (of the form $3\pi$, $\pi\rho$, $\pi\sigma$, and $\pi$ in the three-pion channel),
and distillation together with the GEVP method are employed.
Only states in the rest frame and for a single irrep are considered.
In total, there are 9 (3) three-pion levels and 12 (12) two-pion levels for $M_\pi=300(210)\;$MeV,
with the two-pion levels distributed between the $I=0,1,2$ channels.

\begin{figure}[h!]
\centering
\includegraphics[width=0.9\textwidth]{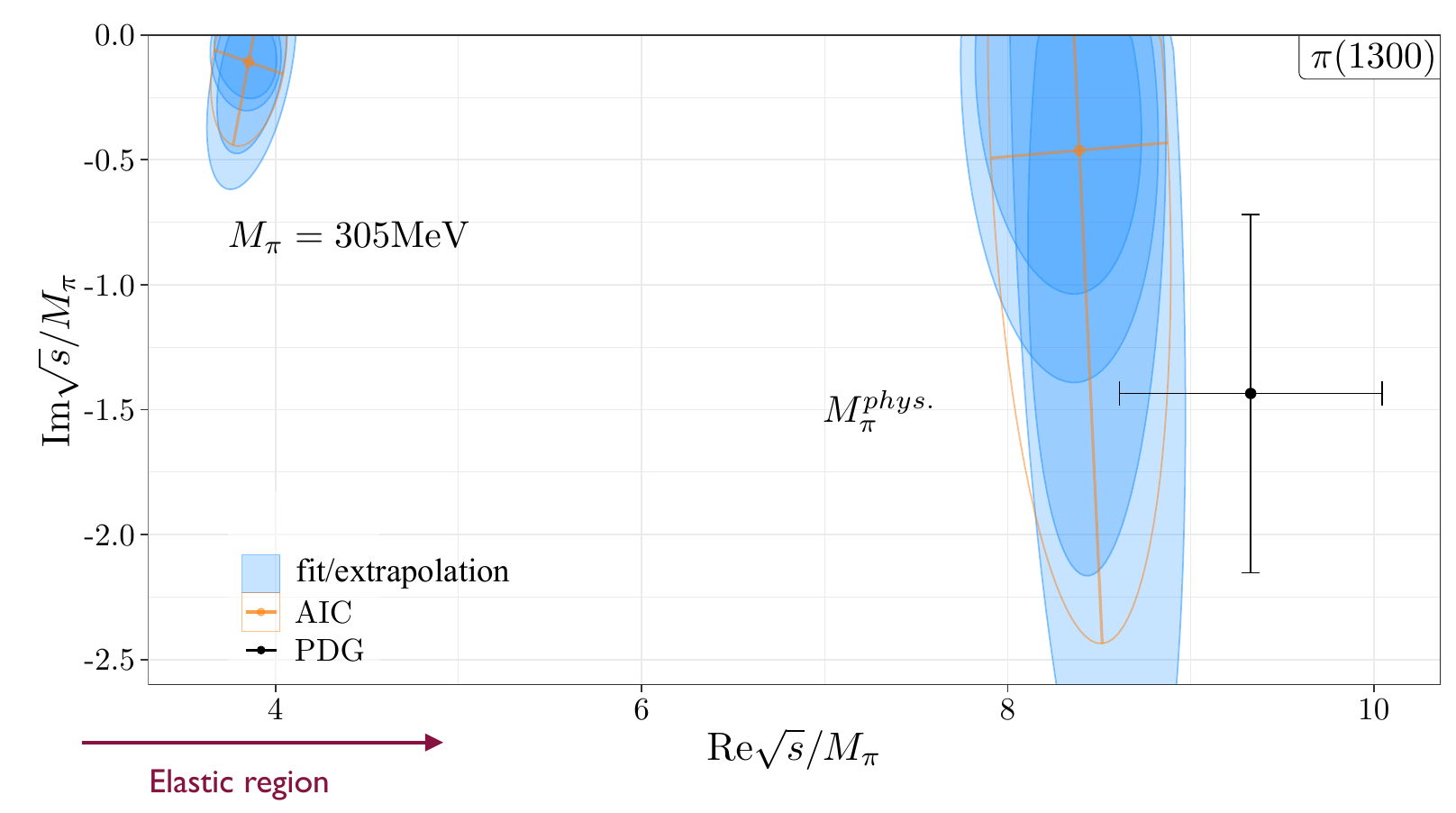}
\caption{Results from Ref.~\cite{Yan:2025mdm} for the resonance pole in the $I=1$, $J^P=0^-$ channel.
Those at $M_\pi=305\;$MeV are obtained from the three-particle formalism alone, while those at the
physical pion mass involve model-based chiral extrapolations. The region in which the three-particle formalism is valid is indicated by the magenta arrow. Shaded areas correspond to $1\sigma$ regions from different fits; the AIC bars are from an average over fits using the Akaike information criterion.}
\label{fig:FVU}
\end{figure}

The calculation is thus of smaller scale in terms of levels than that described in the previous section,
but it is also more challenging because of the large number of quark contractions
that are present for nonmaximal isospin ($I=0$ for $\omega$, and $I=1$ for $\pi(1300)$).
Global fits are performed to two- and three-pion levels, using many choices of K matrix parametrizations,
with that for two-pions based on the inverse-amplitude method, which incorporates constraints from ChPT and
S-matrix theory~\cite{Hanhart:2008mx}. In the best fit, there are 3 two-pion and 6 three-pion parameters, 
and $\chi^2/{\rm dof}=1.08$ for $27$ degrees of freedom. 
The parameters are then inserted into the three-particle integral equations, the solutions to which allow determination of potential resonance poles---see \Cref{fig:FVU}.

In my opinion, the most significant result is that in the top left corner of the plot, 
namely the presence of an excited-pion resonance pole for $M_\pi = 305\;$MeV.
For this pion mass the pole is within the region of validity of the QC3, and its position is determined
essentially by the results from the heavier pion mass alone, for which $M_\pi L \gtrsim 4$.
Thus it involves limited modeling, and shows that the three-particle formalism can be used successfully in the
presence of a resonance.
The extrapolation to physical masses, shown in the right-hand side of the plot, is more model-dependent, 
and also leads to very large errors.

\subsection{Using three-particle methods to study the $T_{cc}^+(3875)$}

The doubly-charmed tetraquark, discovered only in 2021~\cite{LHCb:2021vvq,LHCb:2021auc},
is a manifestly exotic meson with quark composition $cc \bar u \bar d$ and $(I)J^P=(0)1^+$.
It lies just below the $DD^*$ threshold, and decays into $DD\pi$ with a tiny width of about 50\;keV.
Disentangling its structure is an important challenge---Is it primarily a $DD^*$ molecule, or a small tetraquark state, or does it have a large diquark-antidiquark component?---one in which LQCD can play an important role.
For recent reviews, see Refs.~\cite{Chen:2022asf,Hanhart:2025bun}.

\begin{figure}[h!]
\centering
\includegraphics[width=0.4\textwidth]{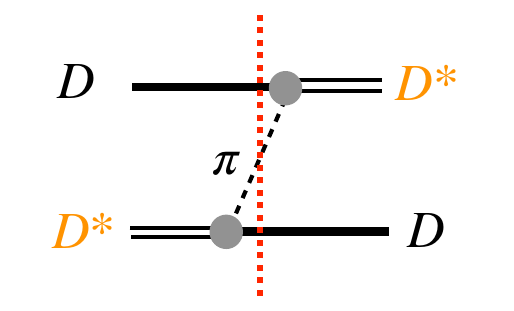}
\includegraphics[width=0.45\textwidth]{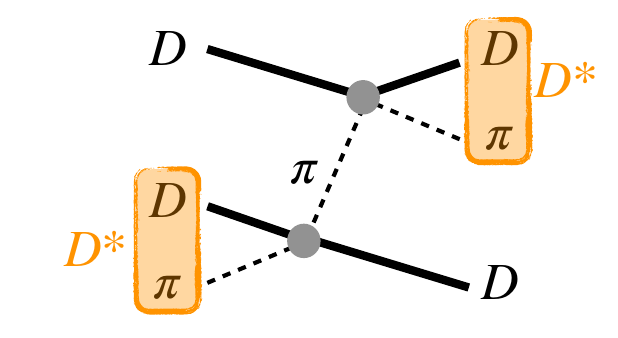}
\caption{Left panel: $u$-channel pion exchange contribution to $DD^*$ scattering, showing three-particle cut.
Right panel: sketch of how $D D^*$ scattering and the $u$-channel exchange emerge in a $DD\pi$ amplitude.}
\label{fig:lhc}
\end{figure}

Given that the $T_{cc}^+$ decays into three hadrons, it would seem to be a natural candidate for the use
of the three-particle formalism.
However, it turns out that one has to move only to slightly higher than physical pion masses for the $D^*$ to become stable to strong decays to $D\pi$, so that the $DD\pi$ and $DD^*$ thresholds interchange, with the latter being the lower of the two. In this situation, the $T_{cc}^+$ can be studied using the two-particle formalism,
and indeed there have been several studies doing so, as described below.
This is not the end of the story, however, because, as was pointed out in Ref.~\cite{Green:2021qol} in the context of two-nucleon interactions, the QC2 breaks down in the vicinity of left-hand cuts.
Such a cut is present in $DD^*$ interactions due to the $u$-channel exchange
 diagram shown in the left panel of \Cref{fig:lhc}, as first stressed in Ref.~\cite{Du:2023hlu}.
If a QC2-based study of $DD^*$ scattering finds a bound state, or virtual bound state, in the vicinity of, or below, the
onset of the left-hand cut, then the results are not valid.
This turned out to be the case for the initial lattice studies~\cite{Padmanath:2022cvl,Whyte:2024ihh}.

A number of resolutions to this problem have been proposed in the context of a two-particle framework:
using a Lippmann-Schwinger equation (LSE) with one-pion exchange (OPE) built in~\cite{Du:2023hlu},
using the HALQCD method~\cite{Lyu:2023xro,Aoki:2025jvi},
generalizing the L\"uscher formalism to explicitly include OPE~\cite{Raposo:2023oru,Raposo:2025dkb},
using the plane-wave version of the QC2 incorporating OPE~\cite{Meng:2023bmz},
and using a finite-volume generalization of the N/D method~\cite{Dawid:2024oey}.
The LSE approach has been applied to the lattice $DD^*$ data in Refs.~\cite{Du:2024wai,Collins:2024sfi,Prelovsek:2025vbr}.
One drawback of these approaches is they break down as the physical point is approached, since this involves
a transition to three-particle decays.

What I wish to briefly describe here is an alternative approach that uses the three-particle formalism~\cite{Hansen:2024ffk}.
The motivation for introducing this (seemingly more complicated) approach is twofold.
First, it is based on a realization that the left-hand cut is a three-particle effect---the exchanged pion goes on shell,
as indicated by the cut in the left-panel of \Cref{fig:lhc}.
Second, using the three-particle formalism allows a smooth transition to the physical situation in which the $D^*$ is unstable and the $T_{cc}^+$ decays to three particles.

The way this approach works is that one considers the $DD\pi$ system, but introduces the $D^*$ as a subthreshold bound-state pole in the $p$-wave $D\pi$ interaction. This pole is certainly present---the only issue is whether it lies above left-hand singularities in the $D\pi$ channel. It turns out that the leading such singularity is due to two-pion exchange, and lies well below the $D^*$ pole. The strength of the $D^*$ pole is determined by the $DD^*\pi$ coupling,
a quantity that must be introduced in all resolutions of the left-hand cut issue, and which can be ultimately determined using lattice calculations. With this set-up, the left-hand cut in the $DD^*$ amplitude due to OPE is automatically
included because $DD\pi$ intermediate states are present. This is illustrated in the right panel of \Cref{fig:lhc}.

The workflow is now as follows. One considers in parallel the $DD\pi$, $DD$ and $D\pi$ systems,
fitting their spectra to, respectively, the QC3 and the $DD/D\pi$ versions of the QC2 (the latter containing the $D^*$ pole). This determines the two- and three-particle K matrices, which are then used in the integral equations to
calculate the $DD\pi$ scattering amplitude. Next, one considers this amplitude in the kinematical configuration in which the $D\pi$ system is below threshold, approaching the $D^*$ pole, and uses the LSZ prescription to extract the $DD^*$ amplitude.
This amplitude can then be studied in the complex plane to search for bound states and resonances.

This rather elaborate-sounding method has been developed in simpler systems in Refs.~\cite{Romero-Lopez:2019qrt,Jackura:2020bsk,Dawid:2023jrj}, 
including
some beautiful work studying the properties of Efimov states~\cite{Dawid:2023kxu}.
It has now been successfully extended to the $T_{cc}^+$ system~\cite{Dawid:2024dgy}, 
and is presently being applied to more extensive
lattice results for the spectra (see the talk by Bai\~ao Raposo).
While it sounds more complicated than the two-particle-based methods noted above, there is little difference in the integral equation part of the calculation.
Indeed, equivalence with both the LSE-based and extended-L\"uscher methods in a simplified setting has
been shown in Ref.~\cite{Dawid:2025wsn}. 
I expect to see extensive results from all methods at next year's lattice conference.

\subsection{The $N\pi\pi$ system}

In this brief section I have two aims: to point out the limitations of the ``LSZ'' method discussed in the previous section,
and to provide a warning about possibly unexpected singularities in three-particle amplitudes.
These results come from Ref.~\cite{\Npp}, in which the RFT formalism for the $N\pi\pi$ system at maximal isospin 
($I=5/2$) is derived.

The initial aim of Ref.~\cite{\Npp} was to extend the RFT formalism so as to allow the study of the Roper resonance,
$N(1440)$. This is the lowest excited baryon with nucleon quantum numbers, and has long been a puzzle in the
quark model, as reviewed, e.g.,  in Ref.~\cite{Burkert:2017djo}.
The Roper decays both to $N\pi$ and $N\pi\pi$, and thus a ``$2+3$'' formalism is needed---one that can treat on-shell cuts associated with both two and three particles.
Such a formalism was derived in Ref.~\cite{\BHSQC}, but only for the case of identical particles.
The initial hope in Ref.~\cite{\Npp} was to finesse the need for this more complicated formalism by using the LSZ method, i.e. by using the three-particle $N\pi\pi$ formalism and obtaining the contributions from $N\pi$ intermediate states by treating the nucleon as a subthreshold pole in the $p$-wave $N\pi$ amplitude.
This approach fails, however, as there is a singularity closer to threshold due to $u$-channel nucleon exchange:
the resulting cut begins at $s=M_N^2+2 M_\pi^2$, which lies above the nucleon pole at $s=M_N^2$.
The QC3 breaks down at the left-hand cut---one would need to consider a four-particle formalism to include the $u$-channel process---so one cannot access the nucleon pole.

Thus it appears that one must use a genuine $2+3$ formalism, and I expect this to be available during the next year.\footnote{%
A $2+3$ EFT-based approach for the Roper resonance in finite volume has previously been presented in Refs.~\cite{Severt:2020jzc,Severt:2022jtg}.}
In the meantime, the $I=5/2$ $N\pi\pi$ system, for which transitions to $N\pi$ are forbidden due to isospin symmetry,
is a useful stepping stone towards a study of the Roper.
This system has nontrivial dynamics, since the $I=3/2$ $N\pi$ channel contains the $\Delta$ resonance,
and in that sense is a baryonic analog of the $3\pi$ system with $I=2$, 
in which the $2\pi$ channels contain both the $\rho$ and $f_0/\sigma$ resonances.
This mesonic system was studied very recently in Ref.~\cite{Briceno:2025yuq},

Turning to the issue of singularities in amplitudes, it was realized in Ref.~\cite{\Npp}
that, for systems with nondegenerate particles, the $3\to3$ Bethe-Salpeter kernel, which plays a central role in the
derivation of the QC3, can have singularities due to four-particle cuts that reduce the allowed parameter space.
This is a rather technical issue that I cannot do justice to here; see Appendix B of Ref.~\cite{\Npp} for a detailed explanation.
I will only summarize the key point, which I think is important for practitioners to be aware of.
This issue arises when one of the nucleons spectates, and has a momentum such that the remaining $N\pi$ pair goes below threshold. In one corner of parameter space, the ``other'' $N\pi$ pair, i.e. that containing the spectator nucleon, can have an invariant mass exceeding $M_N+2 M_\pi$, which leads to a singularity.
Such singularities violate the assumptions of the derivation---they would lead to uncontrolled power-law volume dependence---and the upshot is that one must adjust the cutoff function to avoid them.
One option for so doing is suggested in Ref.~\cite{\Npp}.

\subsection{Dreaming of three neutrons}

Finally, I provide a very brief sketch of the predictions of the QC3 for three neutrons.
In just completed work~\cite{impnnn}, 
Wilder Schaaf and I have addressed the question of what precision will be required
in the determination of $3n$ spectra in order for LQCD to provide significant constraints on the $3n\to 3n$ amplitude.
We have implemented the QC3 derived in Ref.~\cite{\threeneutron}, and determined the spectrum in a set-up that
may be used in future LQCD studies of the $3n$ system: nearly physical  masses with $M_\pi/M_N=0.15$,
and a moderate box size $L=4.3\;$fm such that $M_N L=20$ and $M_\pi L=3$.
We base the two-neutron interactions on experimental results, and keep the two leading terms in the
threshold expansion of $\Kdf$.

\begin{figure}[h!]
\centering
\includegraphics[width=0.7\textwidth]{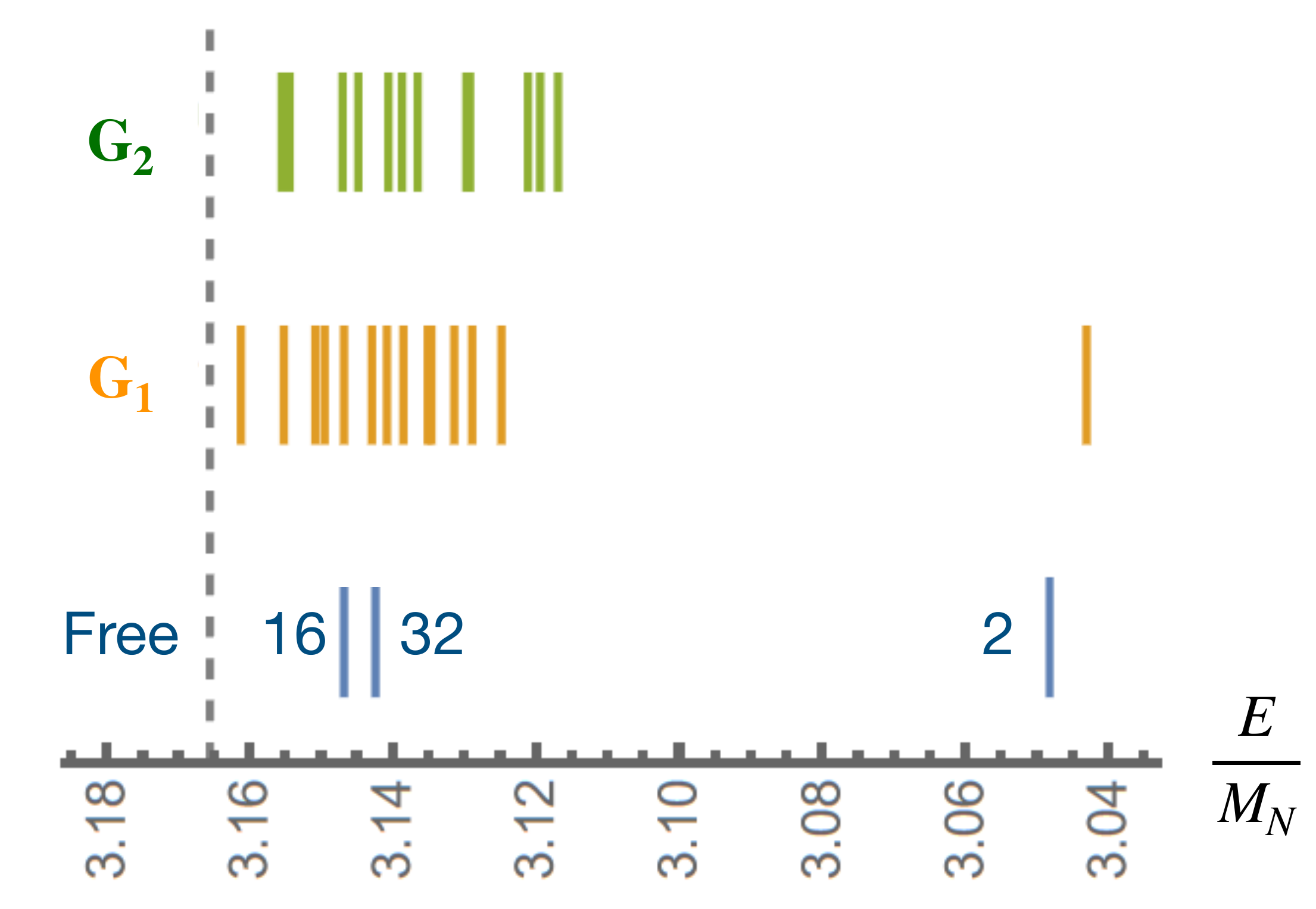}
\caption{Spectrum for states with $\boldsymbol P=(0,0,1) (2\pi/L)$, which live in the two-dimensional irreps $G_1$ and $G_2$.
For comparison, noninteracting levels are shown in the lowest row, along with their degeneracies.
The vertical dashed line shows the inelastic threshold above which the QC3 breaks down.}
\label{fig:nnnspect}
\end{figure}

\begin{figure}[h!]
\centering
\includegraphics[width=0.9\textwidth]{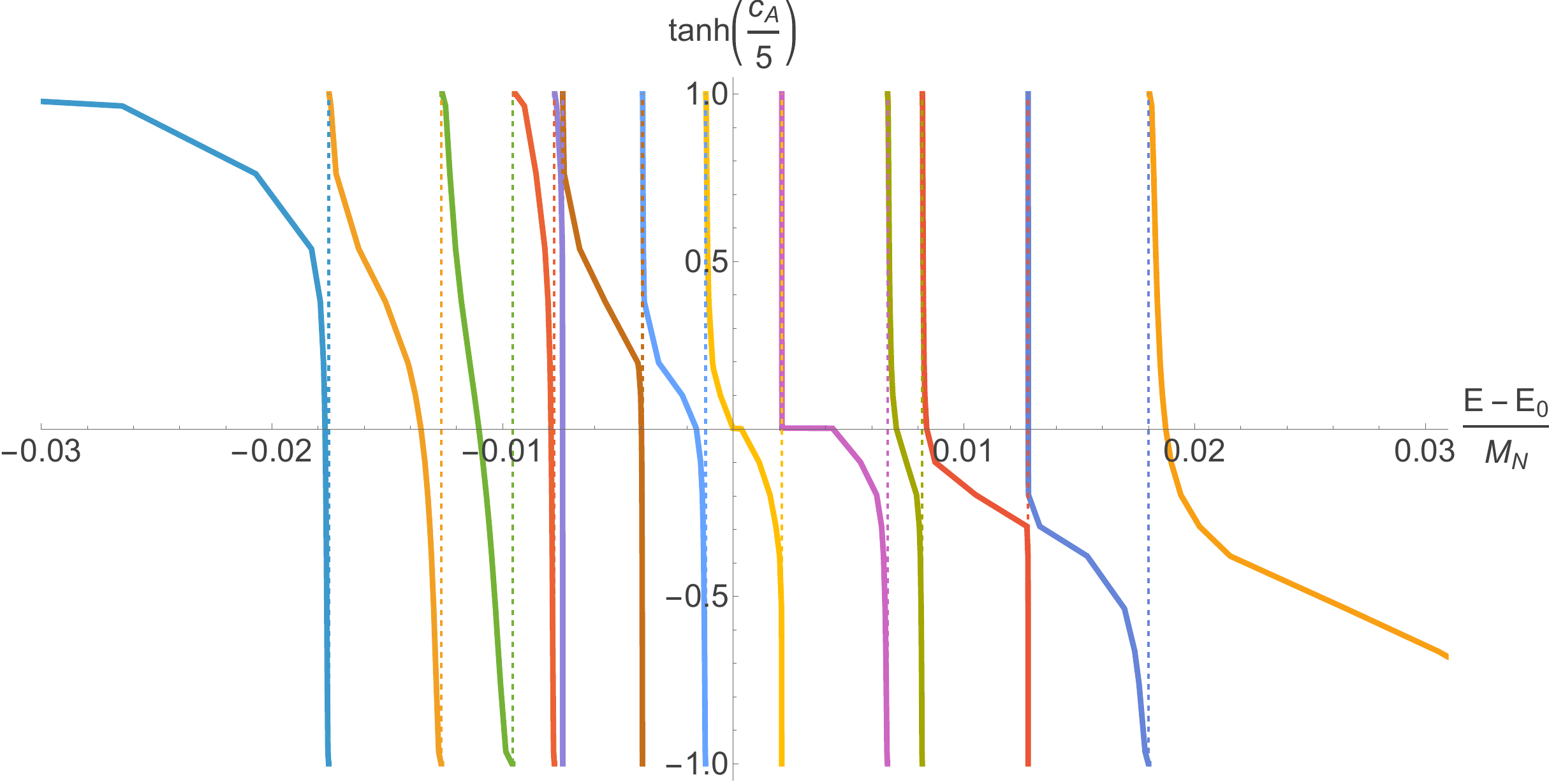}
\caption{Upper band of $G_1$ states for $\boldsymbol P=(0,0,1) (2\pi/L)$, as a function of $\tanh(c_A/5)$, 
where $c_A$ is the coefficient of one of the terms in $\Kdf$.
The energies are measured relative to the lower of the noninteracting energies in this band, at $E_0=3.1424 M_N$.}
\label{fig:nnnspectKdf}
\end{figure}

Examples of the resulting spectra are shown in \Cref{fig:nnnspect,fig:nnnspectKdf}.
In the first figure, the results with $\Kdf=0$ are shown for one frame.
The 50 noninteracting levels, which come in two bands, are spread out by the two-particle interactions.
The spectral density is reduced by separating levels into irreps of the doubled cubic group, here $G_1$ and
$G_2$. Typical spacings range from $5-25\;$MeV.
The pattern of splittings holds information about the form of the two-particle interactions.

In \Cref{fig:nnnspectKdf}, I show what happens to the 13 $G_1$ levels in the upper (left-hand) ``band'' from \Cref{fig:nnnspect} as the coefficient of one of the
terms in $\Kdf$, $c_A$, is varied from large negative to large positive values. The levels shift down by one step, as indicated by the vertical dotted lines. There is much more that can be said about these plots (see Ref.~\cite{impnnn}),
but the key message here is that the impact of $\Kdf$ can be as large as the splittings induced by two-particle interactions. Thus a precision of $\sim 5\;$MeV will allow one to place significant constraints on three-neutron interactions.

\section{Summary and Outlook}\label{sec:outlook}

In the last few years, major progress has been made in the three-particle sector.
The formalism is well established, has been extensively cross-checked, and covers almost all processes of interest.
There are several pioneering end-to-end applications that start with two- and three-particle energy levels and
predict three-particle scattering amplitudes.
I expect that the next few years will see continuing progress in formalism and implementation, 
as well as applications to more challenging systems.
 In particular, the major gaps in the formalism are for $2+3$ systems:
 the $N\pi + N\pi\pi$ system needed for the Roper resonance,
 and the $3\pi + K\bar K$ system needed to access the anomaly-related physics encapsulated in the
 Wess-Zumino-Witten term in ChPT.
 Near-term applications will likely include the $DD\pi$ system at lighter quark masses, so as to study
 the $T_{cc}^+$, and extended calculations of the three-pion system.
 
Although I did not discuss this here, I also note that the formalism has been derived to allow a
calculation of $K\to 3\pi$ decay amplitudes~\cite{Muller:2020wjo,\Kpipipi,Muller:2022oyw}, 
and it would certainly be very interesting (though undoubtably challenging) 
to use this to predict its CP-violating part from LQCD.

There are, however, several important challenges that must be faced in order to obtain physical results
for the most interesting systems.
\begin{itemize}
\item
Reducing model-dependence.
Calculations to date use fairly simple expressions for $\Kdf$.
These may be adequate for weakly-interacting systems such as pseudo-Goldstone bosons at maximal isospin,
but are unlikely to be good enough in the presence of three-particle resonances.
The choice of $\Kdf$ introduces a hard-to-estimate model dependence into the results,
one which is likely to grow in importance as one analytically continues away from the real axis.
Interesting recent work suggests an approach to ameliorate these concerns by using
Bayesian reconstruction together with Nevanlinna interpolation~\cite{Salg:2025now}.

\item
Bringing results to FLAG readiness. 
So far, almost all work is done at a single lattice spacing with at most two box sizes. 
In this regard, it is important to realize that while we need results with all errors controlled,
the errors themselves do not, in many cases, need to be small in order to have an impact.
For example, for the three-nucleon interaction, even a semiquantitative result would be useful.

\item
Comparing formalisms and calculations in detail.
Now that the applications are studying systems with interesting dynamics, it is important to
``look under the hood'' and make sure that the different approaches lead to consistent results for
the scattering amplitudes. There are important differences between the approaches at the technical level.
For example, the FVU approach, as well as the RFT approach adopted in Ref.~\cite{Briceno:2025yuq}, 
use an asymmetric form of the QC3, whereas other RFT calculations use a symmetric form.
The forms used for $\Kdf$ differ also, with Ref.~\cite{Briceno:2024ehy,Jackura:2025wbw,Briceno:2025yuq} 
arguing that an asymmetric form is preferable.
Cutoff functions also differ: the RFT approaches use smooth cutoffs of various types,
while the FVU approach uses a hard cutoff.
There are also differences in the manner in which the final three-particle amplitude is projected onto specific
choices of $J^P$.

I think what is needed here is a workshop where the contributors discuss these differences in detail, and carry out the end-to-end analyses using all approaches on a set of standard datasets: a QC3 bootcamp!

\item
Moving to four or more particles. Doing so is clearly necessary for many applications, e.g. 
studying $\pi(1300)$ for physical masses, a complete treatment of the $\rho$ (since the $4\pi$ decay
is kinematically allowed), and a method for calculating weak decays of $D$ mesons (where decays to
two, three, four and more particles are allowed).
One question here is whether the QC4, QC5, \dots path is feasible.
Aside from the threshold expansion, the four-particle formalism is not yet known,
although some pioneering efforts were described in the talk by Mukherjee.
I can envisage two possibilities.
The optimistic one is that the impact of short-ranged interactions of four or more particles is very small,
and a set of recursive quantization conditions and integral equations involving only $\cK_2$ and $\Kdf$
can be derived and made practical.
The pessimistic one is that the QC4 turns out to be barely practical, while QCN (with $N>4$) is out of reach.

My personal hunch is that the ultimate way forward will involve a clever combination of QC2-4
with inclusive methods such that we are able to effectively squeeze all the information that is possible
out of the lattice data. In this regard, I note that inclusive methods for scattering amplitudes of an arbitrary
number of particles exist in principle, see Ref.~\cite{Bulava:2019kbi}.

\item
Investigate alternative approaches.
Let me mention a couple that I find interesting. Reference~\cite{Patella:2024cto}
proposes to obtain scattering amplitudes from Euclidean correlators using an alternative to the LSZ approach
of Ref.~\cite{Bulava:2019kbi}, namely via Haag-Ruelle theory and approximation formulae.
There is also the three-particle version of the HALQCD approach, which has so far only been
worked out in the nonrelativistic regime~\cite{Doi:2011gq}.

\item
Combine LQCD results with EFT and amplitude-analysis methods.
A wealth of experience in the phenomenology of resonances---normal and exotic---has been accumulated
using inputs from ChPT, dispersion relations, and amplitude analysis.
As LQCD pushes towards comparison with experimental data, it will have to deal with the grungy details
of isospin breaking, electromagnetic effects, triangle singularities, etc,
and it seems very likely to me that this will require learning from and collaborating with the phenomenological
community. The EXOHAD collaboration is one example where such collaboration is being attempted~\cite{EXOHAD}.

\end{itemize}

There is clearly much work to be done, but, for me, this is a very worthwhile and  exciting endeavor,
with the potential to greatly improve our understanding of the strong interactions.

\section*{Acknowledgments}
I thank Andr\'e Bai\~ao Raposo, Ra\'ul Brice\~no, Sebastian Dawid, Jeremy Green, 
Max Hansen, Maxim Mai, Fernando Romero-L\'opez, and Wilder Schaaf for
many discussions on the topics discussed here, and comments on this review.
This work is supported in part by the U.S. Department of Energy grant No. DE-SC0011637, and contributes to the goals of the USDOE ExoHad Topical Collaboration, contract DE-SC0023598.

%

\bibliographystyle{JHEP}
\bibliography{ref}

\end{document}

%% file: shortcuts.tex
\newcommand{\cD}[0]{\mathcal D}

\newcommand{\cK}[0]{\mathcal K}

\newcommand{\cM}[0]{\mathcal M}
\newcommand{\cO}[0]{\mathcal O}

\newcommand{\cS}[0]{\mathcal S}

\newcommand{\df}[0]{\mathrm{df}}

\newcommand{\Kdf}[0]{{\cK_{\df,3}}}

\usepackage{mfirstuc} 
\newcommand{\addReviewer}[2]{
  \expandafter\newcommand\csname #1\endcsname[1]{{\bf \color{#2} \capitalisewords{#1}:\,##1}}
  \expandafter\newcommand\csname #1cor\endcsname[2]{{\color{#2} \capitalisewords{#1}:\,\st{##1}{\bf ##2}}}
  \expandafter\newcommand\csname #1color\endcsname{#2}
}

\definecolor{cardinal}{rgb}{0.77, 0.12, 0.23}
\definecolor{teal}{rgb}{0.0, 0.5, 0.5}

\addReviewer{sebastian}{cardinal}
\addReviewer{fernando}{orange}
\addReviewer{steve}{Maroon}
\addReviewer{MTH}{teal}

\newcommand{\HSQCa}[0]{Hansen:2014eka}
\newcommand{\HSQCb}[0]{Hansen:2015zga}

\newcommand{\HSBS}[0]{Hansen:2016ync}
\newcommand{\HSPT}[0]{Hansen:2015zta}
\newcommand{\HSTH}[0]{Hansen:2016fzj}
\newcommand{\SPT}[0]{Sharpe:2017jej}
\newcommand{\BHSQC}[0]{Briceno:2017tce}
\newcommand{\BHSnum}[0]{Briceno:2018mlh}

\newcommand{\dwave}[0]{Blanton:2019igq}

\newcommand{\largera}[0]{Romero-Lopez:2019qrt}
\newcommand{\BHSSU}[0]{Briceno:2019muc}

\newcommand{\HHanal}[0]{Blanton:2019vdk}
\newcommand{\isospin}[0]{Hansen:2020zhy}
\newcommand{\Kpipipi}[0]{Hansen:2021ofl}

\newcommand{\implement}[0]{Blanton:2021eyf}
\newcommand{\tetraquark}[0]{Hansen:2024ffk}
\newcommand{\DRS}[0]{Dawid:2024dgy}
\newcommand{\threeneutron}[0]{Draper:2023xvu}
\newcommand{\multichannel}[0]{Draper:2024qeh}
\newcommand{\Npp}[0]{Hansen:2025oag}
\newcommand{\BSQC}[0]{Blanton:2020jnm}
\newcommand{\BSequiv}[0]{Blanton:2020gha}
\newcommand{\BSnondegen}[0]{Blanton:2020gmf}
\newcommand{\BStwoplusone}[0]{Blanton:2021mih}

\newcommand{\Akakia}[0]{Hammer:2017uqm}
\newcommand{\Akakib}[0]{Hammer:2017kms}
\newcommand{\AkakiTHa}[0]{Pang:2019dfe}

\newcommand{\MDpi}[0]{Mai:2018djl}
\newcommand{\MD}[0]{Mai:2017bge}

\newcommand{\Akakinum}[0]{Doring:2018xxx}
\newcommand{\Maiaone}[0]{Mai:2021nul}
\newcommand{\HSrev}[0]{Hansen:2019nir}

\newcommand{\MDRrev}[0]{Mai:2021lwb}
\newcommand{\Maiisobar}[0]{Mai:2017vot}

